\documentstyle[12pt,graphicx,color,subfigure,amsmath,cite,ulem]{article}
\newcommand{\comment}[1]{}

\def\issue(#1,#2,#3){{\bf #1}, #2 (#3)}

\def\PREP(#1,#2,#3){Phys.\ Rep. \issue(#1,#2,#3)}
\def\EPJC(#1,#2,#3){Eur.\ Phys.\ J.\ C \issue(#1,#2,#3)}

\def\beq{\begin{equation}}
\def\eeq{\end{equation}}
\def\beqn{\begin{eqnarray}}
\def\eeqn{\end{eqnarray}}
\def\ptslash{\not \! p_T }

\def\mhalf{m_{\frac{1}{2}}}

\def\mygraph#1#2{ \subfigure[]{
   \label{#1}
   \hspace*{-0.6in}                     
   \begin{minipage}[b]{0.5\textwidth}                       
   \centering
   \hspace*{4ex}
   \includegraphics[width=0.9\textwidth,height=0.7\textwidth]{#2}
   \vspace*{-5ex}
   \end{minipage}}
   \vspace*{-1ex}
}

\begin{document}
\begin{center}
  { \Large\bf 
Mixed Neutralino Dark Matter in Nonuniversal Gaugino Mass Models  
\\}
  \vglue 0.5cm
  Utpal Chattopadhyay$^{(a)}$, 
 Debottam Das$^{(a)}$ and D.P.Roy$^{(b)}$
    \vglue 0.2cm
    {\it $^{(a)}$Department of Theoretical Physics, Indian Association 
for the Cultivation of Science, \\2A \& 2B Raja S.C. Mullick Road, Kolkata 700 032, India \\}
 {\it $^{(b)}$Homi Bhabha centre for Science Education, Tata Institute of
   Fundamental Research, \\V.N. Purav Marg,Mumbai-400088,India \\} 
  \end{center}        
\begin{abstract}
We have considered nonuniversal gaugino mass models of
supergravity, arising from a mixture of two superfield
contributions to the gauge kinetic term, belonging to a singlet and a
nonsinglet representation of the GUT group. 
In particular we analyse two models, where the contributing superfields 
belong to the singlet and the 75-dimensional, and the singlet and the 
200-dimensional representations of SU(5).
The resulting lightest superparticle is a mixed bino-higgsino state in the
first case and a mixed bino-wino-higgsino state in the
second. In both cases one obtains 
cosmologically compatible dark matter relic
density over broad regions of the parameter space.
We predict promising signals in direct dark matter detection
experiments as well as in indirect detection experiments
via high energy neutrinos coming from their pair-annihilation in the Sun. 
Besides, we find interesting $\gamma$-ray signal rates that will 
be probed in the Fermi Gamma-ray Space Telescope. We also 
expect promising collider signals at LHC in both cases.
\end{abstract}

\section{Introduction}
\label{intro}
A leading candidate for the cold dark matter of the universe is the lightest 
supersymmetric particle (LSP) of the minimal supersymmetric standard model
(MSSM)\cite{SUSY}. Astrophysical constraints require the 
LSP to be colorless and 
chargeless, while direct dark matter (DM) search 
experiments strongly disfavour 
a sneutrino LSP. Thus the favoured candidate for LSP in the MSSM is the 
lightest neutralino
 \begin{equation}
\label{lsp}
\tilde \chi \equiv \tilde \chi^0_1 = c_1\tilde{B}~+c_2\tilde{W}^3~
+c_3\tilde{H}^0_D~+c_4\tilde{H}^0_U.
\end{equation}
The neutralino mass matrix is given by
\begin{equation}
\label{neutmat}
M_{N}=\left(
\begin{array}{cccc}
M_1 & 0 & -M_Z\cos\beta \sin\theta_W & M_Z\sin\beta \sin\theta_W \\ 0 & M_2 &
M_Z\cos\beta \cos\theta_W   & -M_Z\sin\beta \cos\theta_W  \\ -M_Z\cos\beta
\sin\theta_W & M_Z\cos\beta \cos\theta_W  & 0 & -\mu \\ M_Z\sin\beta 
\sin\theta_W &
-M_Z\sin\beta \cos\theta_W  & -\mu & 0
\end{array} \right),
\end{equation}
where $M_1$,$M_2$ and $\mu$ are the bino,wino and higgsino mass parameters;
and $\tan\beta$ represents the ratio of two Higgs vacuum expectation values. 
The LSP is the lightest eigenstate of the mass matrix. It can be a dominantly 
bino,wino or higgsino state or else a large admixture of these interaction 
eigenstates.

Much of the MSSM phenomenology so far has been done in the context of  minimal
supergravity (mSUGRA)\cite{msugra} model, 
because of its simplicity and economy of 
parameters. This model predicts a dominantly bino LSP over the bulk of the 
parameter space. The bino does not couple to gauge bosons since it does not 
carry any gauge charge. So the main annihilation mechanism of the bino 
dark matter is via $t$-channel
slepton exchange, $\tilde \chi \tilde \chi \xrightarrow{\tilde l} 
{l^+  l^- }$. However the large slepton mass limits from 
LEP\cite{Amsler:2008zzb}, 
particularly in the mSUGRA model, makes this annihilation mechanism 
inefficient. This leads to a gross overabundance of the DM relic density 
over the bulk of the mSUGRA parameter space. There are only narrow strips 
of parameter space, like the stau coannihilation
($\tilde \chi \tilde \tau \xrightarrow{\tau} 
{\tau \gamma }$) and resonant annihilation ($\tilde \chi \tilde \chi \xrightarrow
{A} {b \bar b, t \bar t, \tau^+ \tau^- }$) regions giving cosmologically 
compatible DM relic density\cite{WMAPdata} which require however 
some stringent mass correlations between the annihilating LSPs and the 
intermediate particle in the $s$-channel resonance   
or between a coannihilating sparticle and the LSP.   

In contrast, the higgsino and wino carry weak isospin $I=\frac{1}{2}$ and $1$
respectively. So a higgsino and wino-dominated LSP can pair-annihilate 
efficiently via their couplings to the $W/Z$ gauge bosons, leading to an 
underabundance of DM relic density for sub-TeV LSP masses. One can get 
cosmologically compatible relic density only for 
relatively large LSP masses of 
$M_{\tilde H}\simeq 1~{\rm TeV}$ and $M_{\tilde W}\simeq 
2~{\rm TeV}$, which make
these models inaccessible to LHC. 
They can only be probed at a multi-TeV 
linear collider like CLIC\cite{CLIC,CLICworks}.

It is evident from the above discussion that a mixed bino-higgsino or 
bino-wino LSP is  expected to give cosmologically compatible DM relic 
density for sub-TeV LSP masses, which can be probed at the LHC. 
This has been described as the `well-tempered' scenario 
in Ref.\cite{Arkani-Hamed:2006mb}.  The 
DM relic density and the detection phenomenology of this scenario 
have been studied in a model independent way in 
Ref.\cite{masiero_mixedlsp}. It should be mentioned here that a 
very important example of the mixed bino-higgsino LSP occurs in the 
so-called hyperbolic branch/focus point 
(HB/FP)\cite{hyper,focus,focusLHC1,focusLHC2} region of mSUGRA model. 
This is a narrow strip at the edge of the mSUGRA parameter space, 
corresponding to very large (multi-TeV) scalar masses.

In this work, we investigate a class of nonuniversal gaugino mass models, 
where the mixed neutralino LSP can be realised in a simple and economical way.
 In particular we construct two models $-$ the first giving a mixed 
bino-higgsino LSP and the second giving a mixed bino-wino LSP. 
The latter model also gives a triply 
mixed bino-wino-higgsino LSP over a significant part of the parameter space. 
For each of these two models we find cosmologically compatible DM relic 
density over broad bands passing through the middle of the parameter space 
in the $m_0-\mhalf$ plane. 
We also investigate the expected signals in both cases for direct and indirect 
DM detection experiments. In each case we find promising signals for future 
direct detection experiments. Likewise we find promising indirect detection 
signals in each case for the IceCube experiment, in the form of high energy 
neutrinos coming from DM
pair-annihilation inside the Sun. We also estimate the corresponding line and 
continuum $\gamma$ ray signals coming from their pair-annihilation in the 
galactic core.
We conclude with a brief discussion of the expected collider signatures 
of these models at LHC.

\section{Nonuniversal Gaugino Mass Models for mixed neutralino DM:}
\label{nugmformixedneut}
SUGRA models with nonuniversal gaugino masses at the GUT scale have been 
studied in many earlier 
works\cite{etcEllis:1985jn,NathMixedRep,Chattopadhyay:2001mj,
nugminter,ucdphiggsino,kingrobertsdp,nugmvarious,Huitu:2008sa,choietc}. 
We shall only summarize the 
main results here, focusing on the simplest and most predictive GUT group,
 $SU(5)$. Here the  gauge kinetic function that is related to 
the GUT scale gaugino masses, 
arises from the vacuum expectation value of the $F$-term of a chiral superfield 
$\Phi$, which is responsible for 
SUSY breaking. This results in a dimension five term in the lagrangian,
\beqn
L \supset {\frac{{<F_\Phi>}_{ij}}{M_{Planck}}} \lambda_i \lambda_j,
\eeqn  
where $\lambda_{1,2,3}$ are the $U(1)$, $SU(2)$ and $SU(3)$ gaugino fields 
$-$ bino ($\tilde B$), wino ($\tilde W$) and gluino 
($\tilde g$). Since the gauginos belong to the adjoint representation of 
$SU(5)$, $\Phi$ and $F_\Phi$ can belong to any of the irreducible 
representations appearing in their symmetric product, i.e. 
\beqn
{(24 \times 24)}_{symm} =1+24+75+200 . 
\eeqn
The minimal SUGRA model assumes $\Phi$ to be a singlet, 
implying equal gaugino masses at the GUT scale. On the other hand if 
$\Phi$ belongs to one of the nonsinglet representations of $SU(5)$, 
then these gaugino masses are unequal 
but related to one another via the representation invariants.  Thus the 
three gaugino masses at the GUT scale in a given representation $n$ 
are determined in terms of a single SUSY breaking mass parameter 
$m^n_{1/2}$ by 
\beqn
M^{G}_{1,2,3} = C^n_{1,2,3} m^n_{1/2}, 
\label{relativegauginos}
\eeqn
where 
\beqn
\label{wts}
C^{1}_{1,2,3}=(1,1,1), ~C^{24}_{1,2,3}=(-1,-3,2), 
~C^{75}_{1,2,3}=(-5,3,1), ~C^{200}_{1,2,3}=(10,2,1).
\eeqn

  	The nonuniversal gaugino mass models are known to be consistent 
with the observed universality of the gauge couplings at the GUT 
scale~\cite{etcEllis:1985jn,Chattopadhyay:2001mj}, with
$\alpha^G (\simeq 1/25)$. Since the gaugino masses evolve like the 
corresponding gauge couplings at one-loop level 
of the renormalisation group equations (RGE), the 
three gaugino masses at the electroweak (EW) 
scale are proportional to the corresponding gauge couplings, i.e.
\beqn
M_1 & = & (\alpha_1/\alpha_G) M_1^G  \simeq  (25/60) C_1^n 
m^n_{1/2} \nonumber \\
M_2 & = & (\alpha_2/\alpha_G) M_2^G \simeq  (25/30)C_2^n 
m^n_{1/2} \nonumber \\ 
M_3 & = & (\alpha_3/\alpha_G) M_3^G  \simeq  (25/9) C_3^n m^n_{1/2}.  
\label{gauginoEW}
\eeqn
For simplicity we shall assume a universal SUSY breaking scalar mass $m_0$ 
at the GUT scale. Then the scalar masses at the electroweak scale are 
given by the renormalisation group evolution 
formulae~\cite{Ibanez:1984-85,carenakomine}.
  A very important SUSY breaking mass parameter at this scale is 
$m^2_{H_U}$, which appears in the EW symmetry breaking condition, 
\beqn
\mu^2 + M_Z^2/2= {{m_{H_D}^2 -m_{H_U}^2\tan^2 \beta} \over {\tan^2\beta -1} }
\simeq -m_{H_U}^2.
\eeqn
 The last equality holds for the $\tan\beta \geq 5$  region, which is 
favoured by the Higgs mass limit from LEP2~\cite{hlim}.  Expressing 
$m_{H_U}^2$ 
at the right in terms of the GUT scale parameters via the one-loop RGE
gives~\cite{carenakomine}
\beqn
\mu^2 + \frac{1}{2} M_Z^2  & \simeq &  
-0.1m_0^2 +2.1 {M_3^G}^2 -0.22 {M_2^G}^2 
-0.006{M_1^G}^2 +0.006 M_1^G M_2^G +  \nonumber \\ 
& & 0.19 M_2^G M_3^G + 0.03 M_1^G M_3^G,
\label{musqeqn}
\eeqn
at a representative value of $\tan\beta=10$, neglecting the contribution 
from the trilinear coupling term at the GUT scale. Moreover, the coefficients 
vary rather mildly over the moderate $\tan\beta$ region.  

Although we 
shall use exact numerical 
solutions to the two-loop RGE in our analysis, one important result is 
worth noting from the simple formulae of 
Eqs.~(\ref{relativegauginos},\ref{wts},\ref{gauginoEW},\ref{musqeqn}). 
For the singlet and $24$-plate representations of the SUSY breaking 
chiral superfield $\Phi$, we have $|M_1| < |M_2|,|\mu|$. This 
corresponds to a bino LSP, resulting in overabundance of DM relic density 
over most of the parameter space. On the other hand, for the $75$-plate 
and $200$-plate representations we get
$|\mu|<|M_1|,|M_2|$.  This corresponds to a higgsino LSP, resulting in 
underabundance of DM relic density\cite{ucdphiggsino}.

In general the gauge kinetic function may contain several chiral superfields, 
belonging to different representations of $SU(5)$, which gives us the 
freedom to vary the relative magnitudes of $M_1$,$M_2$ and $\mu$ continuously. 
In particular we shall consider two models, where the gauge kinetic function 
contains a mixture of singlet plus $75$-plet superfields~($1+75$) 
in the first case and a 
 singlet plus $200$-plet superfields~($1+200$) in the second. 
By adjusting the relative contribution from the two superfields we shall 
obtain a mixed bino-higgsino LSP in the first case and a 
mixed bino-wino LSP in the second, each giving favourable DM relic density 
over large regions of the parameter space. 
It should be mentioned here that 
mixtures of singlet and nonsinglet superfield contributions 
were first discussed in Ref.\cite{etcEllis:1985jn}. 
Moreover, 
the $(1+24)$, $(1+75)$ and $(1+200)$ models were considered in the 
context of i) direct detection of neutralino DM in Ref.\cite{NathMixedRep},  
and ii) a bino DM in Ref\cite{kingrobertsdp},  
where in the latter the relative contribution from the two 
superfields were adjusted to recover the bulk annihilation region of 
the right DM relic density in each case. Nonuniversal gaugino mass models 
based on such mixed representations have also been considered more recently 
to investigate the Higgs production phenomenology via SUSY cascade 
decay\cite{Huitu:2008sa}. To the best of our knowledge however, 
this is the first investigation of such nonuniversal gaugino mass models, 
to obtain the right relic density via mixed neutralino DM. As in 
\cite{kingrobertsdp}, we shall vary the relative contribution 
of the two superfields 
via a single parameter, i.e 
\beqn
m^{1}_{1/2}=(1-\alpha_{75})m_{1/2} \ \ \ \rm and \ \ \ m^{75}_{1/2}=\alpha_{75}m_{1/2}, 
\eeqn
for the (1+75) model and 
\beqn
m^{1}_{1/2}=(1-\alpha_{200})m_{1/2}\ \ \ \rm and \ \ \ m^{200}_{1/2}=\alpha_
{200}m_{1/2},
\eeqn 
for the (1+200) model. 
Then one sees from Eqs. (\ref{relativegauginos},\ref{wts},\ref{gauginoEW},\ref{musqeqn}) that for the (1+75) model,
\beqn
\alpha_{75}\equiv 0.50 \Longrightarrow |M_1| \simeq |\mu| \simeq m_{1/2} <|M_2|
\eeqn  
for $m_0 \geq m_{1/2}$. 
This leads to a mixed bino-higgsino LSP. One also sees from 
Eq.~(\ref{musqeqn}) that decreasing $m_0$ leads to increase of $|\mu|
$ and hence decrease of higgsino fraction of LSP. Moreover one sees from eqns   (\ref{relativegauginos},\ref{wts},\ref{gauginoEW}) that for (1+200) model 
\beqn
\alpha_{200}\equiv 0.1 \Longrightarrow |M_1| \simeq |M_2|,
\eeqn  
leading to a mixed bino-wino LSP. Again one sees from equation  
(\ref{musqeqn}) that increasing $m_0$ leads to decrease of $|\mu|$, 
resulting in
a triply mixed bino-wino-higgsino LSP in the large $m_0$ region. We shall 
present the results for the $(1+75)$ and $(1+200)$ models in the following 
sections.

\begin{table}[!hb]
\begin{center}
\begin{tabular}{|c|c|c||c|c|c|c|}
\hline
$\alpha_{75}$ &$|M_1|$&$\mu$  & $\alpha_{200}$ &$M_1$&$M_2$&$\mu$\\
\hline
 - & (GeV)& (GeV) & - &(GeV) & (GeV) & (GeV) \\
\hline
0.3   &177  &552  &0.1   &  400 & 442  & 612 \\
\hline
0.4   &307  &519  &0.12 &  439  &451 &611  \\
\hline
0.45   &372  & 495  &0.15   &  496 &  463  &609  \\
\hline
0.475 & 404  &484  &0.2   & 592  &483  & 604 \\
\hline
0.5   &437 & 473  &0.3   & 783  & 523  &593 \\
\hline
0.55   &502  &448  & 0.45   & 1073 & 584 & 568 \\
\hline
\end{tabular}
\caption{
Variation of mass parameters with 
$\alpha_{75}$ 
(for the (1+75) model) and $\alpha_{200}$ (for the (1+200) 
model) for $m_0=m_{1/2}=500$~GeV, $A_0=0$,$\tan\beta=10$, 
and sign$(\mu)=+$ve.
} 
\label{xtratab}
\end{center}
\end{table}

We have probed how the nature of the LSP varies with the choice of 
mixing parameters in Table~\ref{xtratab}.
We show $M_1$ and $\mu$ for a few values of 
$\alpha_{75}$ for the (1+75) model for a given set of other 
parameters. 
Similarly we show $M_1$, $M_2$ and $\mu$ 
for a few values of $\alpha_{200}$ for the (1+200) model.
One can infer the dominant component
of the LSP from the relative values of these masses.
In the (1+75) model,
the LSP is dominantly bino for $\alpha_{75} \leq 0.45$, leading to a generic
overabundance of DM. On the other hand it is dominantly higgsino for 
$\alpha_{75} \geq 0.55$, leading to a generic underabundance 
of DM. In the (1+200) model,
the LSP is dominantly wino for $\alpha_{200} \geq 0.15$, 
becoming dominantly
higgsino for $\alpha_{200} \geq 0.45$. 
There is a generic underabundance
of DM throughout the wino/higgsino dominated 
LSP region of $\alpha_{200} \geq 0.15$.
On the other hand, $\alpha_{200}<0.1$ results in 
a dominant bino content of the LSP leading to a generic
overabundance of DM. 

\noindent
Thus we have found the optimised mixing parameters to be 
\beqn
\label{mixing}
\alpha_{75}= 0.475 \ \ \ \rm and  \ \ \ \alpha_{200}= 0.12
\eeqn
so as to get the most favourable DM relic density. 
There 
is admittedly some 
finer adjustment involved in these parameters so that  
we may be able to probe such large mixing zones of LSP.  
This is analogous to the delicate correlations 
between the mass parameters existing 
in the stau coannihilation or the resonant annihilation regions 
that provide with cosmologically favourable DM relic density 
in mSUGRA models, as pointed out in Section~\ref{intro}.

Note however that once the mixing parameter is so adjusted, then
the $(1+75)$ and $(1+200)$ models are each as predictive as the mSUGRA model. 
Note that in the first case we need roughly equal contributions from the 
singlet and the $75$-plet superfields to achieve a 
favourable DM relic density, 
while in the second case it is achieved through a dominantly singlet 
contribution with only
$\sim 10\%$ admixture from the $200$-plet superfield.

\section{DM Relic Density for the $(1+75)$ Model:}
\label{Relicfor1plus75}
We have computed the DM relic density for the $(1+75)$ and $(1+200)$ models 
in terms of the GUT scale mass parameters $m_0$ and $m_{1/2}$ 
using SUSPECT\cite{suspect} for the numerical evaluation of the 
two-loop RGEs. The  
DM relic density was evaluated using micrOMEGAs\cite{micromegas} 
and cross-checked with DARKSUSY\cite{darksusy}.

\begin{figure}[!h]
\centering
\includegraphics[width=0.7\textwidth,height=0.5\textwidth]{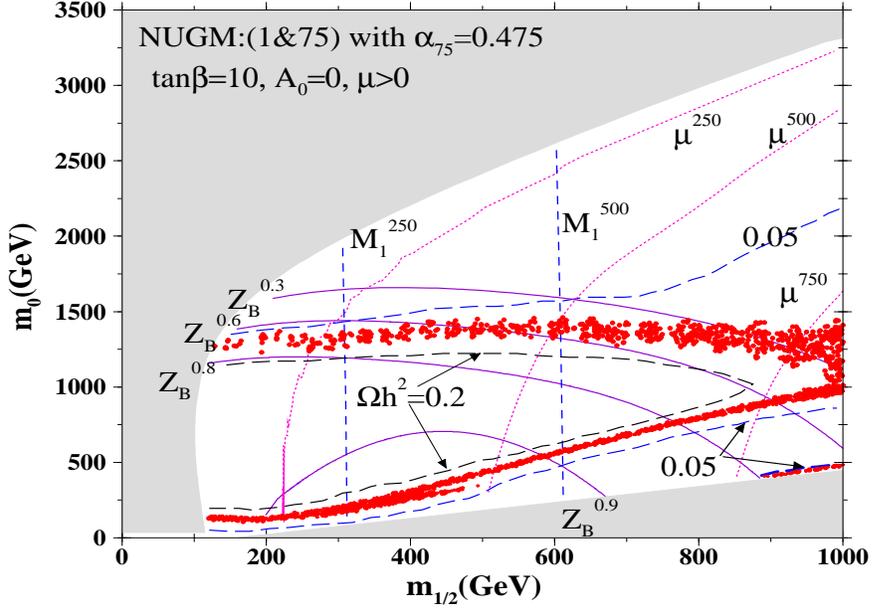}
\caption{Allowed and disallowed zones 
of the $(1+75)$ model. The WMAP DM relic density satisfying 
regions are shown in red in 
the $m_0-m_{1/2}$ plane for $\tan\beta=10$, $A_0=0$, 
$\mu>0$ corresponding to $\alpha_{75}=0.475$ of Eq.\ref{mixing}.  The upper 
disallowed region (in gray) corresponds to no EWSB ($\mu^2 < 0$). The lower 
gray region corresponds to stau becoming the LSP. Contours are drawn 
for $M_1$, $\mu$, $Z_B$ and $\Omega_{CDM}h^2$. 
}
\label{75-spectrum}
\end{figure}

Fig.\ref{75-spectrum} shows the DM relic density in the $m_0-m_{1/2}$ plane 
for the $(1+75)$ model with the mixing parameter of eq.(\ref{mixing}) 
and a moderate value of $\tan\beta=10$. For simplicity we have set $A_0=0$ 
and positive $\mu$. The upper disallowed region (in gray) 
corresponds to the lighter chargino being smaller 
than the LEP limit\cite{Amsler:2008zzb} or no EWSB 
($\mu^2 < 0$); and the lower one to a stau LSP. The allowed region 
is mapped by constant $M_1$ and $\mu$ contours. At the intersection points 
$M_1=\mu$, the LSP is an equi-mixture of bino and higgsino 
($Z_B \equiv c_1^2=0.5$). The bino fraction increases with decreasing 
$m_0$, as indicated by the fixed $Z_B$ contours. The region satisfying the 
DM relic density value of the WMAP data\cite{WMAPdata} 
\begin{equation}
\Omega_{CDM}h^2 = 0.1099 \ \ \pm 0.0186~(3\sigma),  
\label{relicdensity}
\end{equation}
is indicated by the bands of red dots. The contours of fixed 
$\Omega_{CDM}h^2 = 0.05 \ \ \rm and~0.2$ are also shown on the two sides 
of these red bands.

We see from Fig.\ref{75-spectrum} that almost half the parameter space 
corresponds to DM relic density being in the right ball-park of 
$\Omega_{CDM}h^2 = 0.05-0.2$.  And even the precise WMAP value of 
Eq.(\ref{relicdensity}) is satisfied by two fairly
thick bands passing through the middle of the parameter space. The upper band 
corresponds to the DM being a roughly equal admixture 
of $\tilde B - \tilde H$, 
where their pair-annihilation occurs mainly via the gauge coupling of the 
higgsino component. With decreasing $m_0$ the bino component ($Z_B$) 
increases, leading to the increase of DM relic density upto 
$\Omega_{CDM}h^2 \simeq 0.6$. However further decrease of $m_0$ leads to 
resonant pair-annihilation via the pseudo-scalar higgs boson $A$, as 
$2 M_1 \rightarrow m_A $.
This is the dominant dark matter annihilation 
mechanism for the lower 
red band, satisfying the WMAP relic density of 
Eq.~(\ref{relicdensity}). The region below this band has an underabundance of 
DM relic density because of the resonant pair annihilation as 
$m_A \simeq 2 M_1$. 
This is the so-called funnel region. The lower edge of the funnel is 
marked by the red strip at the bottom right, adjacent to the lower boundary. 
We have found this strip to be dominated by resonant pair annihilation of 
DM via $A$, with only 
$\sim 10\%$ contribution from the stau coannihilation. 
The $\Omega_{CDM}h^2 \simeq 0.05$ 
contour (its lower branch) practically overlaps with the 
upper edge of this red strip.

\begin{table}[!vt]
\begin{tabular}[vt]{cllllll}
\hline
parameter & A & B & C & D & E & F \\
\hline
$m_{1/2}$ &300  &600 &800 &300  &600 &800\\
$m_0$ &1325  &1400 &1340 &185  &550 &800\\
$\mu$ &268 & 496 &  654 &323 & 560 &  694 \\
$M_1$ &242 & 493 &  662 &237 & 489 &  660 \\
$m_{\tilde\chi^0_{1}}$  &227& 468 &  631 &231 &481 &  645 \\
$m_{\tilde\chi^0_{2}}$ &254&489&649&302 &552&689 \\
$m_{\tilde\chi^0_{3}}$&289&524&687&334&571&711 \\
$m_{\tilde\chi^0_{4}}$&501&970&1285&490&957&1277\\
$m_{{\tilde\chi_1}^{+},{\tilde\chi_2}^{+}}$ & 255,501 &490,970  
& 649,1285 &304,491 &552,957 &689,1277 \\
$m_{\tilde g}$ & 794 & 1437 & 1845 &725 &1381&1809 \\
$m_{\tilde t_1}$ & 903 & 1232 & 1435 & 485&965&1280  \\
$m_{{\tilde t_2},{\tilde b_1}}$ & 1256,1246 &1704,1694  
& 1993,1983 &727,651 &1367,1288 &1800,1712 \\
$m_{\tilde l}$ & 1300$-$1400 &  1400$-$1600 & 1440$-$1680 &280$-$440
&680$-$950 &960$-$1300 \\
$m_{\tilde q_{1,2}}$ & 1400$-$1500 &  1800$-$1940 & 2000$-$2220 &650$-$750
&1300$-$1480 &1725$-$1965 \\
$m_A(\simeq m_{H^+},m_H)$ & 1382 &  1648 &  1786 &534&1091&1460  \\
\hline
\end{tabular}
\caption{MSSM masses in~GeV for a few sample parameter points 
for the $1+75$ model.}
\label{tab75}
\end{table}

Table~\ref{tab75} lists the SUSY spectra for three representative points on 
each of the two main branches, satisfying DM relic density from 
WMAP(Eq.\ref{relicdensity}). 
Note that the SUSY spectra show an inverted mass hierarchy , where the 
lighter stop($\tilde t_1$) is much smaller than the other squark masses. 
This is related to the large negative contribution from the 
terms associated with top-Yukawa coupling in the renormalisation 
group evolution of the third generation of squark masses\cite{focusLHC1}. 
This feature is more pronounced for the 
upper branch, represented by the first three columns
due to larger values of $m_0$. But it is quite 
significant for the lower branch as well. 
Note also an approximate  
degeneracy among the three lighter 
neutralinos ($\chi^0_{1,2,3}$) and the lighter chargino 
(${\tilde \chi}^{+}_1$). This is again more prominent for the upper branch,
which corresponds to a larger mixing between higgsinos and bino.
Consequently,  
the coannihilation of $\chi^0_{1}$ with $\chi^0_{2}$ and 
${\tilde \chi}^{\pm}_1$ make important contributions to the annihilation 
process. The dominant annihilation processes for the upper branch are
\beqn
\label{75-up}
\chi^0_{1}\chi^0_{1},~\chi^0_{1}\chi^0_{2},~\chi^0_{1}{\tilde \chi}^{\pm}_1 
\ \ \ \rightarrow \ \ \ WW,~ZZ,~f\bar f ~
(\rm both~via ~s~and~t- \rm channel~processes )
\eeqn
which are driven by the gauge coupling of the higgsino component. 
On the other hand the lower branch is dominated by the 
resonant annihilation processes 
\beqn
\label{75-down}
\chi^0_{1}\chi^0_{1},~\chi^0_{1}\chi^0_{2},~\chi^0_{1}{\tilde \chi}^{+}_1 
\ \ \ \rightarrow \ \ \ t\bar t,~b\bar b,\tau \bar \tau,~ t \bar b,~\tau
\nu_\tau ~(\rm via ~s-\rm channel ~A,H,H^+)
.
\eeqn

\noindent
Since the gauge coupling of $W/Z$ to the LSP pair goes like the square of its 
higgsino component, the annihilation process of Eq.\ref{75-up} are strongly
suppressed with the decrease of the higgsino component to $\sim 10\%$ for 
the lower branch.
On the other hand the $A$ coupling to the LSP pair goes like the product 
of its higgsino and gaugino components and hence remains significant for a 
higgsino component of $\sim 10\%$. Even more importantly there is a large 
resonance enhancement for this region as $2 m_{\chi^0_{1}} \rightarrow m_A $.
This also includes some amount of coannihilation of $\chi^0_{1}$ and 
$\chi^+_{1}$ in the $s$-channel. 

Note finally that a higgsino component of $\sim 10\%$ in the LSP is 
still large compared to the mSUGRA model except a few regions 
like HB/FP zones. This is why one can get a 
resonant annihilation region here even for a moderate value of 
$\tan\beta$\cite{ucddNUSM08}. 

\section{ Direct \& Indirect Detection rates for the $(1+75)$ Model:}
\label{detectionsfor1plus75}
We have estimated the DM signals for the $(1+75)$ model for direct and 
indirect detection experiments using the DARKSUSY\cite{darksusy}. The direct 
detection signal is based on the elastic scattering of the DM on a heavy 
nucleus like Germanium or Xenon. The main contribution comes from the 
spin-independent interaction of the DM $\chi_1^0$ with nucleon, which adds 
coherently in the nucleus. This is dominated by the Higgs exchange. 
Fig.\ref{xsectionA} shows the scatter plot of this cross-section against 
the DM mass. The discovery limits of present and proposed direct detection 
experiments are also shown for comparison. 
We have shown the limits from CDMS (Ge) 2005\cite{cdmsGe2005},
XENON-10\cite{xenon102007} and future SuperCDMS (Snolab)\cite{supercdmssno}
and XENON1T\cite{xenon1t} experiments in the figure.
The upper and lower 
shaded (red) branches correspond respectively to the upper and lower WMAP 
relic density satisfying branches of Fig.\ref{75-spectrum}. Since the Higgs 
coupling to the neutralino DM is proportional to the product of its higgsino 
and gaugino components, one gets a larger cross-section for the upper branch 
of Fig.\ref{xsectionA}, where DM corresponds to a roughly equal 
admixture of bino 
and higgsino. This comes within the detection limit of the proposed 
SuperCDMS experiment\cite{supercdmssno}. Both the branches will be 
covered by the 
proposed 1-Ton Xenon\cite{xenon1t} experiment. 
Note that these cross-sections are larger 
than those of the mSUGRA model except the focus point region.

\begin{figure}[!htb]
\vspace*{-0.05in}
\mygraph{xsectionA}{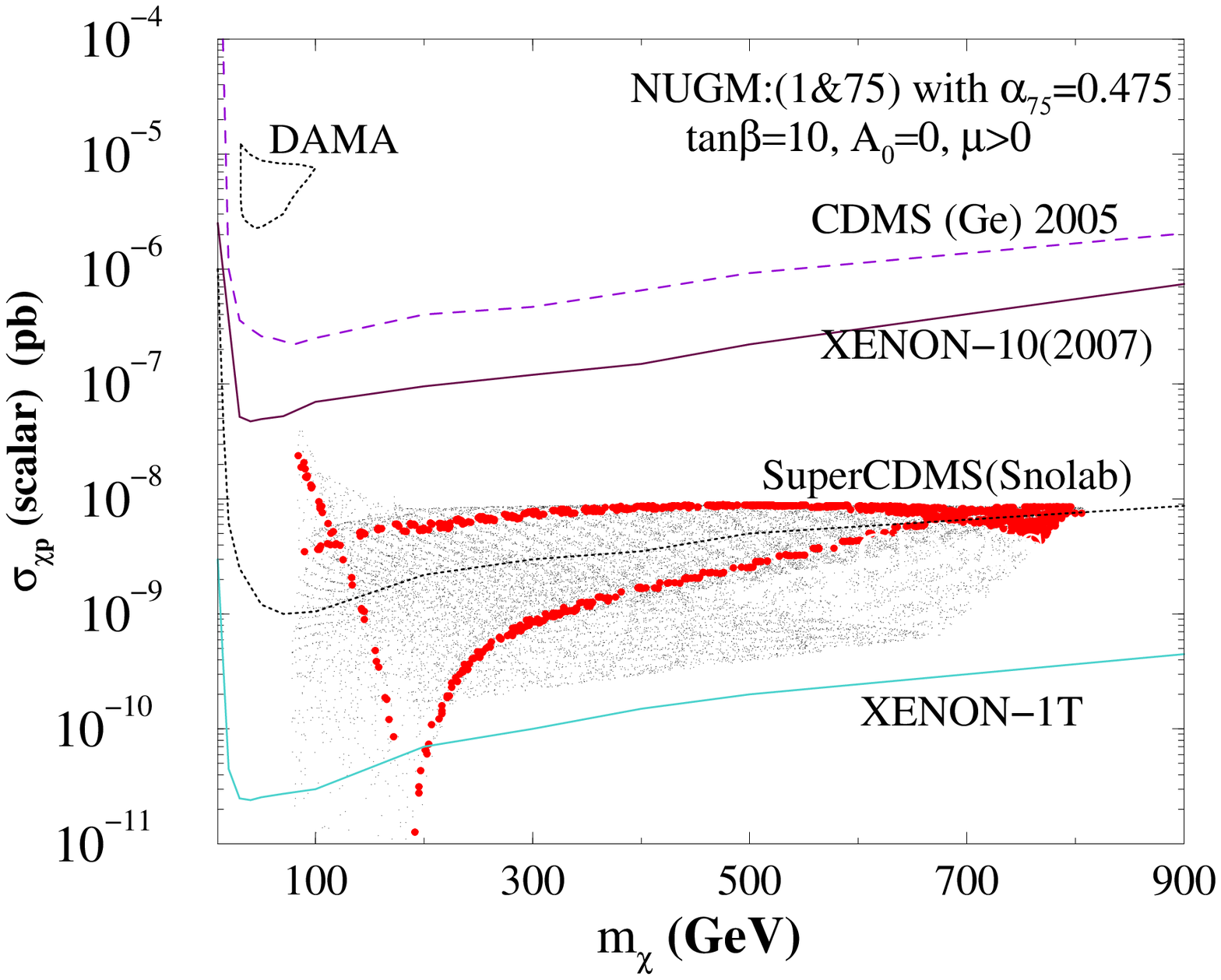}
\hspace*{0.5in}
\mygraph{xsectionB}{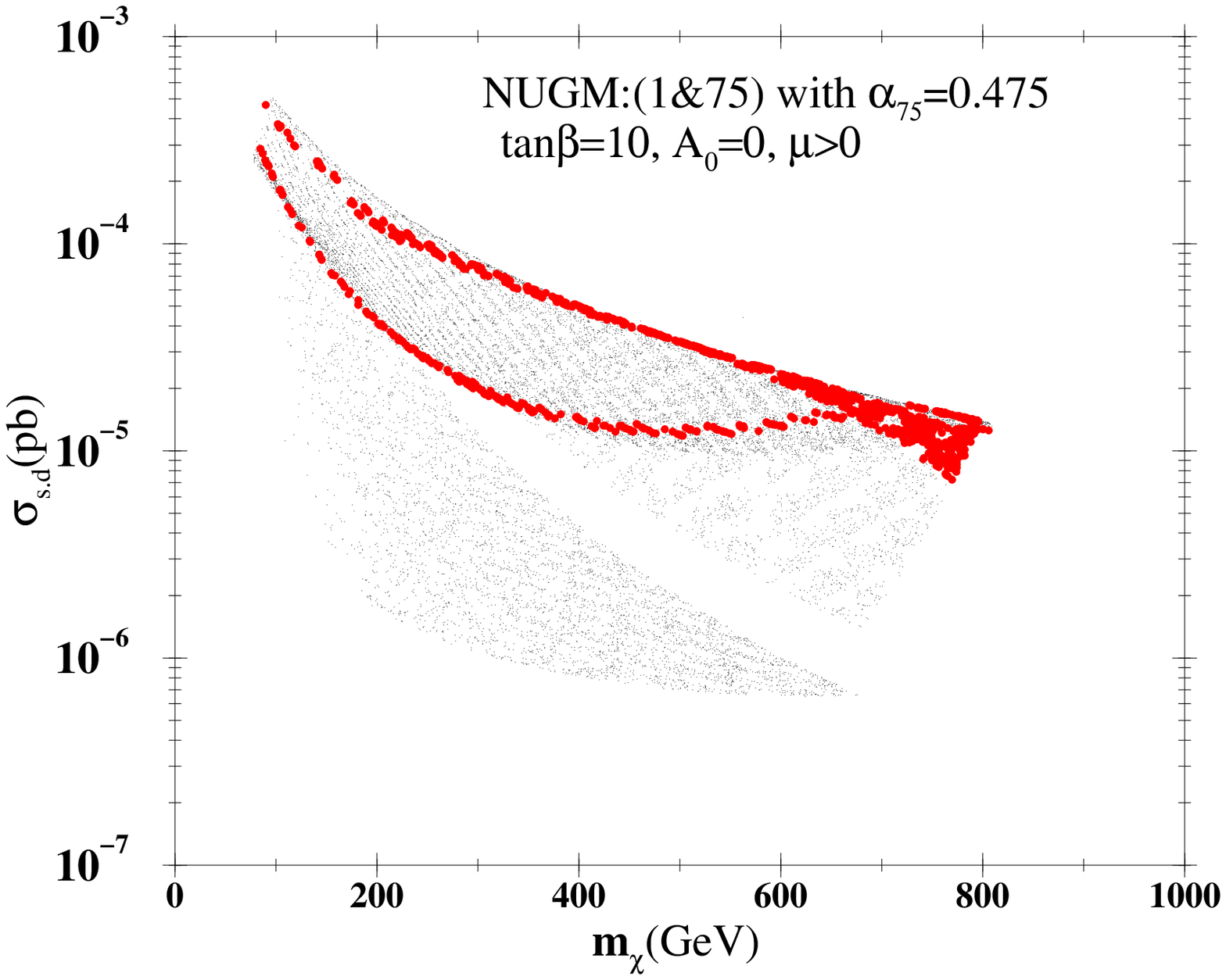}
\caption{(a): Spin-independent neutralino-proton cross section for 
the $(1+75)$ model vs. LSP mass for the parameters shown in 
Fig.\ref{75-spectrum}. The WMAP satisfied regions are shown in red. 
Various limits from the recent (DAMA, CDMS and XENON-10) and 
the future (SuperCDMS and XENON-1T) experiments are shown. 
(b): Spin-dependent neutralino-proton cross section for 
the $(1+75)$ model vs. LSP mass for the parameters shown in 
Fig.\ref{75-spectrum}. The WMAP satisfied regions are shown in red. 
}
\label{xsection75}
\end{figure}

Fig.\ref{xsectionB} shows the corresponding scatter plot of the spin-dependent 
cross-section, which is dominated by $Z$ exchange. Again upper and 
lower shaded (red) branches correspond to WMAP satisfying branches of 
Fig.\ref{75-spectrum} respectively. Since the $Z$ coupling to the 
neutralino DM is proportional to the square of its higgsino component, 
one gets a larger cross-section for the upper branch of Fig.\ref{75-spectrum}, 
corresponding to a larger higgsino component of DM. The spin-dependent 
cross-sections in this model are significantly larger than those of the 
mSUGRA model (except the focus point region). But still they are much below 
the detection limits of any present or proposed direct detection 
experiments\cite{directdetexpt}.

\begin{figure}[!h]
\centering
\includegraphics[width=0.7\textwidth,height=0.5\textwidth]{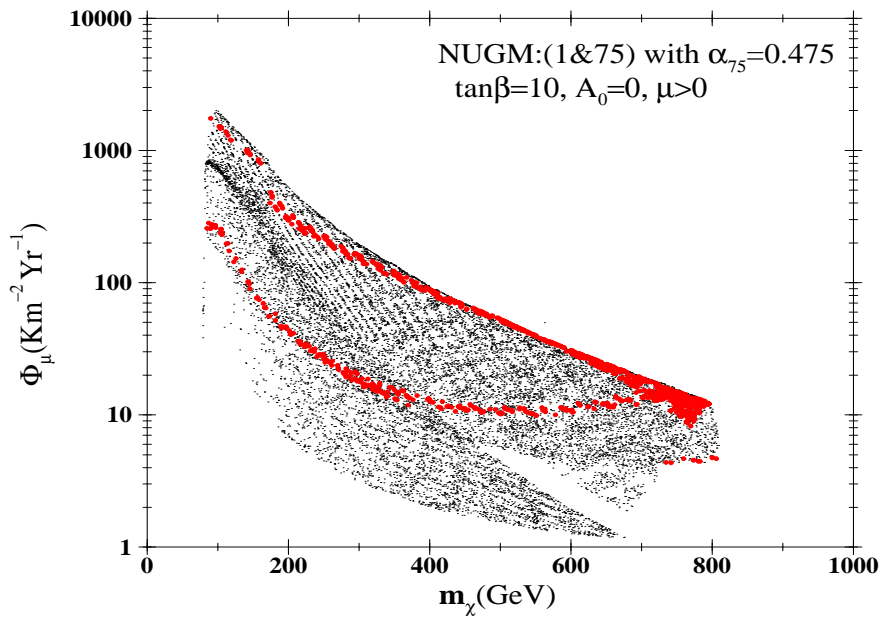}
\caption{Neutralino annihilation induced muon flux 
from the Sun in ${km}^{-2} {yr}^{-1}$ vs 
LSP mass for the $(1+75)$ model for the parameters shown in
Fig.\ref{75-spectrum}. The WMAP satisfied regions are shown in red.
}
\label{muflux}
\end{figure}

A very promising indirect detection experiment for DM is via the high 
energy neutrinos coming from their pair-annihilation in the solar core. 
Since the annihilation rate at equilibrium is balanced by the DM capture 
rate inside the Sun, the resulting neutrino signal is proportional to the 
$\chi_1^0-p$ cross-section. This is dominated by the above-mentioned 
spin-dependent interaction via $Z$ exchange.
Fig.\ref{muflux} shows the model prediction for the rate of 
muon signal events, resulting from these high energy 
neutrinos, in a ${\rm km}^2$ size 
neutrino telescope like
IceCube\cite{Abbasi:2009kq}. 
Again upper and lower shaded (red) branches correspond to 
the upper and lower WMAP satisfying branches of Fig.\ref{75-spectrum}. 
One can even see a small
red strip at the right end, corresponding to a similar one in 
Fig.\ref{75-spectrum}. One sees a very promising signal with $\geq 10$ 
events/year at the IceCube. Since the $Z$ coupling to the neutralino DM 
goes like the square of its higgsino component, the size of this signal 
is much larger than that of the bino dominated DM of the mSUGRA model.

We have also computed the high energy $\gamma$-ray signal coming from the pair 
annihilation of the DM at the galactic core, assuming the standard 
NFW profile 
of DM distribution near the galactic core\cite{NFW}. Fig.\ref{linegamma} 
shows the signal rate of line $\gamma$-rays coming from the DM pair 
annihilation, $\tilde \chi
 \tilde \chi \rightarrow \gamma \gamma (\gamma Z)$ with $E_{\gamma}\simeq 
m_\chi$ 
via $W$-boson loop. Again this decay amplitude is proportional to the square 
of the higgsino component of the neutralino DM. Consequently  the upper and 
lower shaded (red) branches correspond to the respective WMAP satisfying 
branches of Fig.\ref{75-spectrum}, while the small red strip at the right 
end correspond to the corresponding one of Fig.\ref{75-spectrum}. 

\begin{figure}[!htb]
\vspace*{-0.05in}
\mygraph{linegamma}{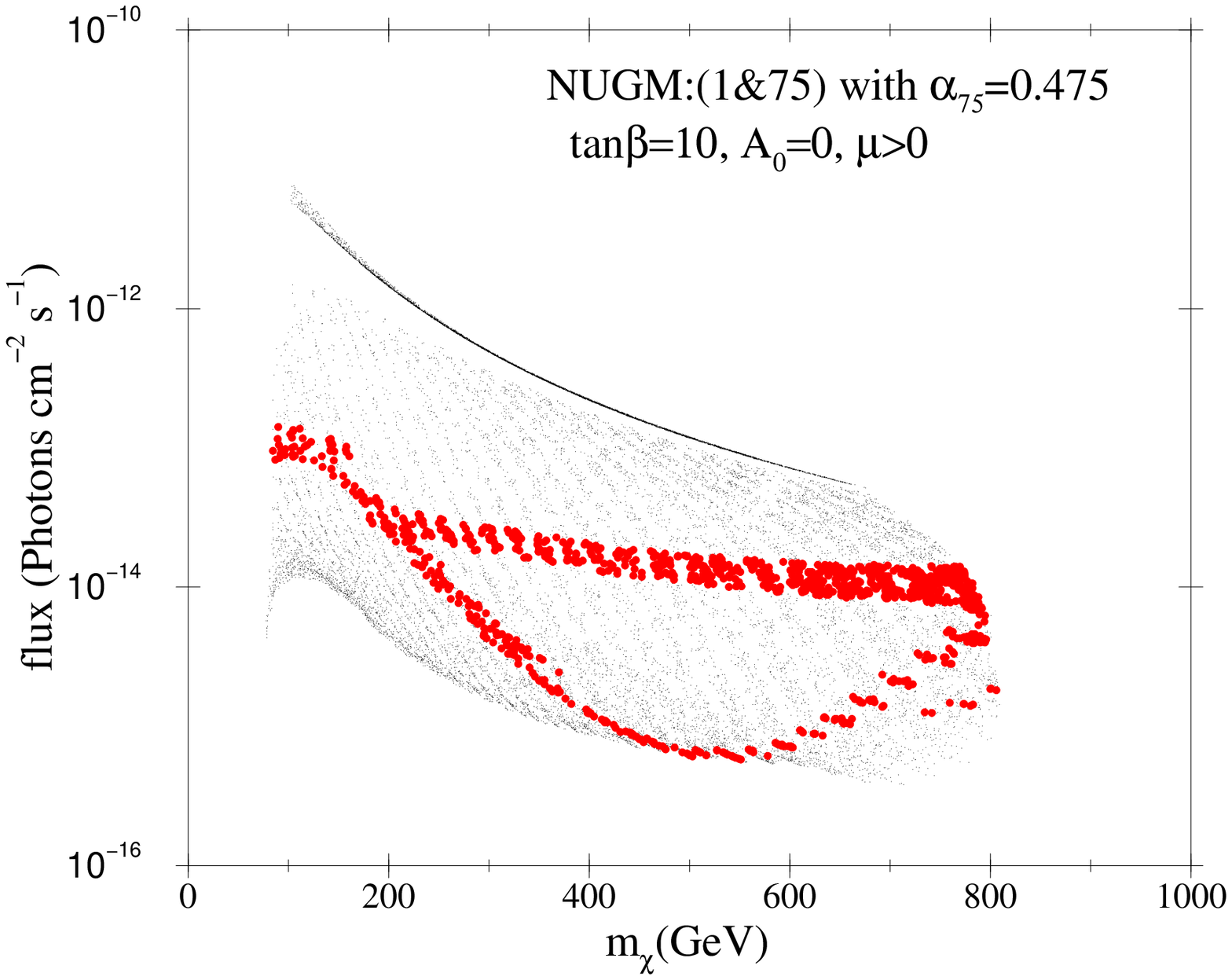}
\hspace*{0.5in}
\mygraph{contgamma}{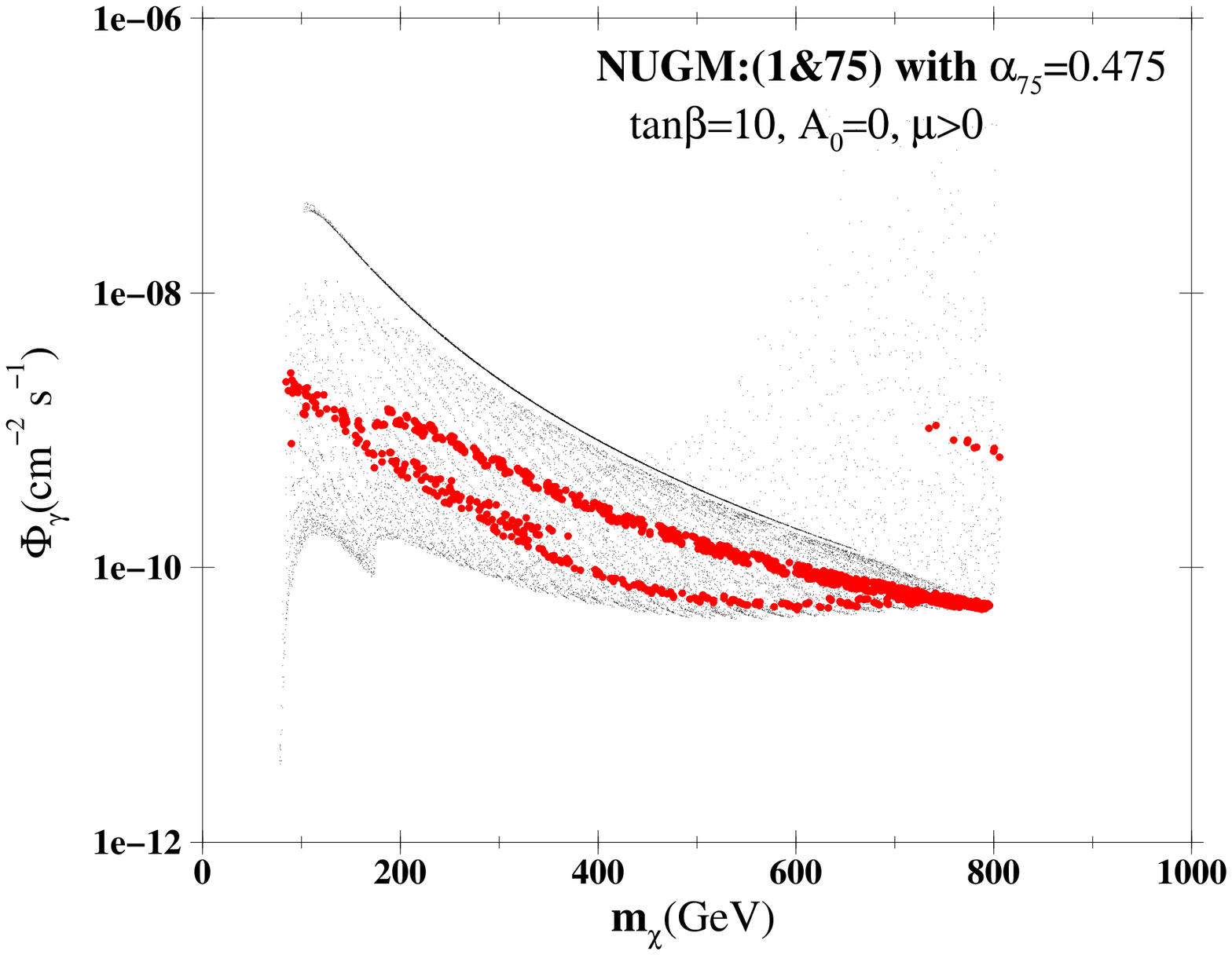}

\vspace*{0.5cm}
\hspace*{1.5in}
\mygraph{75phenergy}{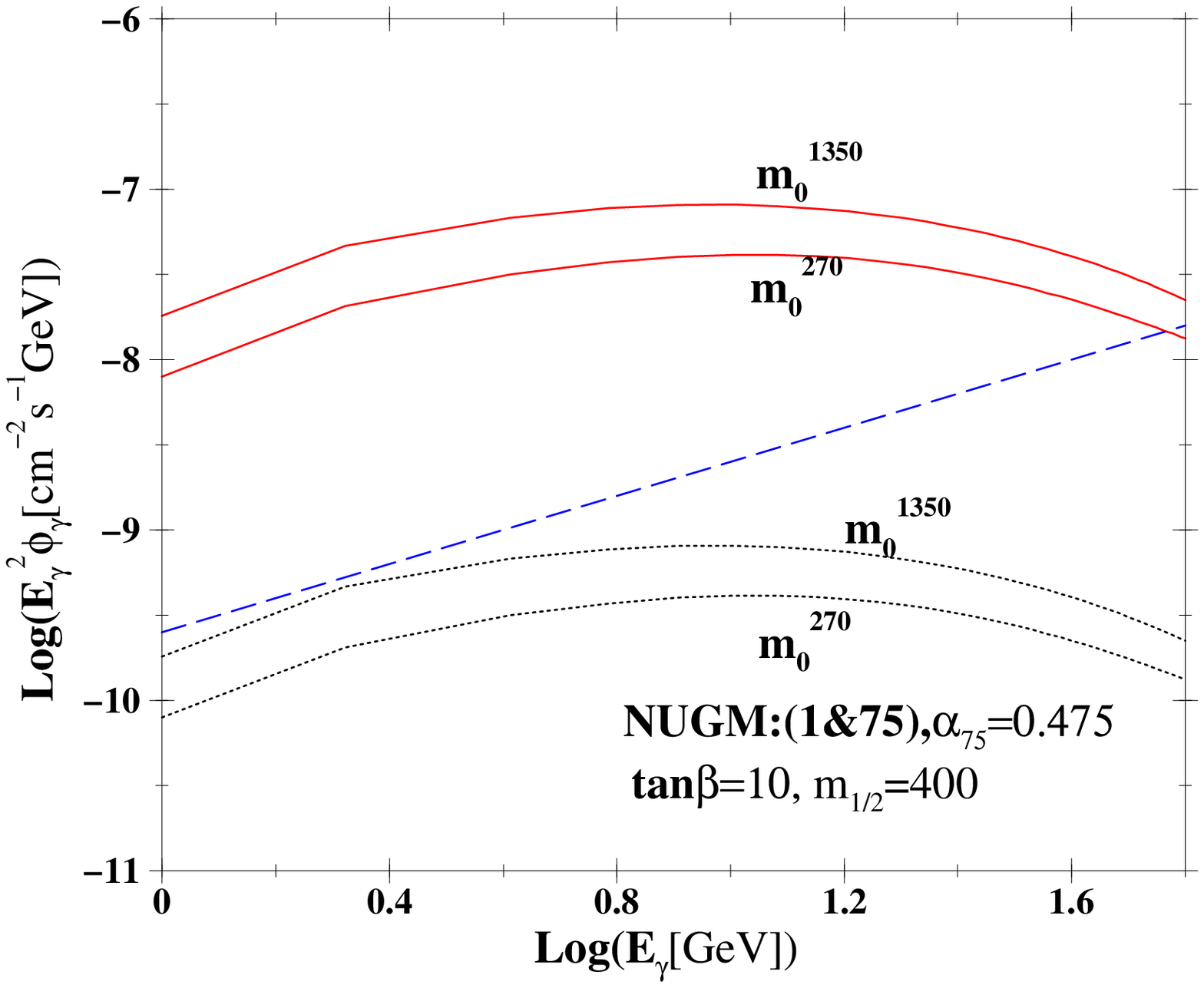}

\caption{
(a):Monochromatic $\gamma$-ray flux from DM 
pair annihilation near the galactic center 
vs. the LSP mass 
shown for the NFW profile of DM halo distribution and an aperture size of 
$\Delta \Omega = 10^{-3}$~sr for the $(1+75)$ Model. 
The parameters are same as in 
Fig.\ref{75-spectrum}. The WMAP satisfied regions are shown in red.
(b): Same as (a) except that the vertical axis refers to a continuous 
$\gamma$-ray flux above a threshold energy of $1$ GeV. (c):
Photon spectra expected from two particular LSPs corresponding to 
$m_0=270$~GeV and $m_0=1350$~GeV for the $(1+75)$ Model with 
$\mhalf=400$~GeV for an NFW profile. 
The other parameters are same as in Fig.\ref{75-spectrum}. 
The upper set of curves corresponds to a boost\cite{boostref} 
factor of $10^2$. The 
discovery limit of FGST/GLAST is shown as a blue straight line.  
}
\label{75-photon}
\end{figure}

 Fig.\ref{contgamma} shows the signal rate of continuum  $\gamma$-rays 
coming from the tree level pair
annihilation processes of Eq.\ref{75-up} and  Eq.\ref{75-down}. 
Production of neutral pions and their subsequent decays into photons
is the most important mechanism for this signal\cite{Cesarini}. While  
the $b \bar b$ channel is the most prominent channel for the 
continuum  $\gamma$-rays signal, the $WW$ and $ZZ$ channels are also very 
significant. The two shaded (red) bands of Fig.\ref{contgamma} correspond 
to the two WMAP relic density satisfying bands of
Fig.\ref{75-spectrum}. The isolated red strip on the top right corresponds
to that on the bottom right of Fig.\ref{75-spectrum}, which is dominated by 
the $b \bar b$ channel.
Finally Fig.\ref{75phenergy} shows the 
continuum $\gamma$-ray spectrum for some representative points of this model, 
along with the discovery limit
of the Fermi Gamma-ray Space Telescope (FGST) experiment or formerly known as 
GLAST\cite{glast1,glast2}. One would need a 
boost\footnote{
Processes like adiabatic compression 
may cause a large enhancement of the DM density near the galactic center 
(see e.g.\cite{boostref}). Larger DM annihilation rate 
may also result from existence of clumps within the halos of 
galaxies\cite{Carr:2006cw}. 
A boost factor essentially  
estimates a few such effects.} 
factor
of 
$\sim 10^2$ to see this $\gamma$-ray signal in FGST/GLAST. Note however that 
one can raise (lower) the signal rate by a factor of $\sim 10^3$ by assuming a 
more spiked (flat) DM profile like that of Moore profile\cite{Moore} (or a spherically 
symmetric isothermal core profile\cite{isothermalcore}).

\section{DM Relic Density for the $(1+200)$ Model:}
\label{Relicfor1plus200}
\begin{figure}[!h]
\centering
\includegraphics[width=0.7\textwidth,height=0.5\textwidth]
{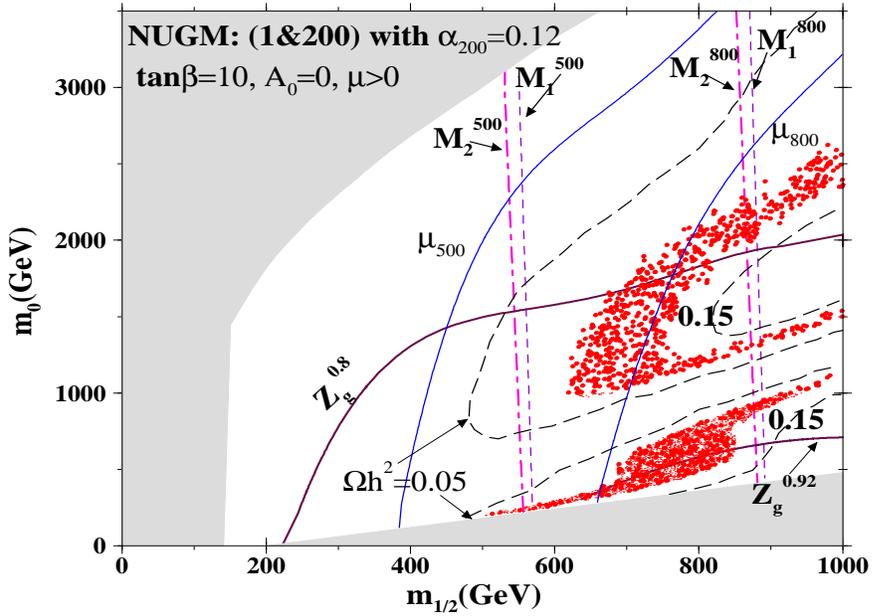}
\caption{
Allowed and disallowed zones of the 
$(1+200)$ model. The WMAP DM relic density satisfying 
regions are shown in red in the 
$m_0-m_{1/2}$ plane for $\tan\beta=10$, $A_0=0$, 
$\mu>0$ corresponding to $\alpha_{200}=0.12$ 
of Eq.\ref{mixing}.  The upper 
disallowed region (in gray) corresponds to no EWSB ($\mu^2 < 0$). The lower 
gray region corresponds to stau becoming the LSP. Contours are drawn 
for $M_1$, $M_2$, $\mu$, $Z_g$ and $\Omega_{CDM}h^2$. 
}
\label{200-spectrum}
\end{figure}

Fig.\ref{200-spectrum} shows the DM relic density in the $m_0-m_{1/2}$ 
plane for the $(1+200)$ model with the mixing parameter of eq.(\ref{mixing}). 
Again the upper and 
lower disallowed regions corresponds no EWSB ($\mu^2 < 0$); and to a stau LSP
respectively.  The allowed region is mapped by constant $M_1$ and $M_2$ 
along with the constant $\mu$ contours. We see from these contours that the 
LSP is an 
equimixture of bino and wino ($\tilde B$-$\tilde W$), along with a significant 
higgsino component in the $m_0 > 1$~TeV region. In fact the intersection 
points of these contours ($M_1,~M_2,~\mu = 500 ~\rm and~ 800 ~GeV$) mark the 
region of equally mixed $\tilde B$-$\tilde W$-$\tilde H$ LSP. The combined 
gaugino component of the LSP is indicated by the fixed 
$Z_g (\equiv c_1^2+c_2^2)$ contours. This, when 
subtracted from unity would be a measure of the higgsino 
component.  The region satisfying the DM relic density value of 
Eq.\ref{relicdensity} from WMAP is indicated by the bands of red dots. 
We see two thick bands in the middle of the parameter space. Note that these  
DM relic density satisfying bands correspond to LSP mass $\geq 500$ GeV. 
The contours of fixed $\Omega_{CDM}h^2 = 0.05~\rm and~0.15$ are also shown on 
the two sides of these red bands. The bulk of the parameter space outside 
these red bands corresponds to underabundance of DM relic density. 
In particular the corridor
cutting through the two bands correspond to underabundance due to rapid 
resonant annihilation of DM pair via $A$-boson (i.e. the funnel region). 
For a clearer
understanding of this model we list the SUSY spectra for two representative 
points from each of the two DM relic density satisfying branches 
in Table~\ref{tab200}.
This model requires a rather heavy SUSY spectrum for satisfying the DM 
relic density constraint. Note that the SUSY spectra of this model show an 
inverted mass hierarchy like the previous model, where the lighter 
stop ($\tilde t_1$) is significantly lighter than the other squarks. 
Note also the close degeneracy among the lighter neutralino and 
chargino states ($\chi^0_{1,2}~\rm and~\rm 
{\tilde \chi}^{+}_1$) due to large bino-wino mixing. Therefore the 
coannihilation of $\chi^0_{1}$ with $\chi^0_{2}$ and ${\tilde \chi}^{\pm}_1$ 
make important contributions to the annihilation process as in the previous 
case. Both the branches are dominated by the annihilation process of 
Eq.\ref{75-up}, driven by the gauge coupling of the wino 
(along with that of higgsino at large $m_0$). But there is a $30-40\%$ 
contribution coming from the resonant annihilation processes 
of Eq.\ref{75-down} on the strips adjacent to the 'funnel' corridor, where
$2 m_{\chi^0_{1}} \rightarrow m_A $. Note that the higgsino component of 
the LSP in this region is $\geq 10\%$, which is large enough for resonant 
annihilation via $A$ even for a moderate value of $\tan\beta~(=10)$.

\begin{table}[!vt]
\begin{tabular}[vt]{cllll}
\hline
parameter & A & B & C & D  \\
\hline
$m_{1/2}$ &725  &900 &725 &900 \\
$m_0$ &1450 &1357 &590 &950\\
$\mu$ &792 & 983 & 846 &1009 \\
$M_1$ &652 & 814 & 646 &811\\
$m_{\tilde\chi^0_{1}}$  &633& 798 &  633 &797\\
$m_{\tilde\chi^0_{2}}$ &657&818&650&815 \\
$m_{\tilde\chi^0_{3}}$&794&985&848&1011 \\
$m_{\tilde\chi^0_{4}}$&822&1009&869&1032\\
$m_{{\tilde\chi_1}^{+},{\tilde\chi_2}^{+}}$ & 643,818 &807,1005  
& 641,865 &806,1028 \\
$m_{\tilde g}$ & 1700 & 2045 & 1637 &2017 \\
$m_{\tilde t_1}$ & 1460 & 1649 & 1216 & 1540 \\
$m_{{\tilde t_2},{\tilde b_1}}$ & 1813,1801 &2013,2001  
& 1478,1452 &1860,1843  \\
$m_{\tilde l}$ & 1535$-$1555 &  1505$-$1530 & 800$-$830 &1160$-$1190
\\
$m_{\tilde q_{1,2}}$ & 2000$-$2050 &  2170$-$2240 & 1520$-$1600 &1950
$-$2040 \\
$m_A(\simeq m_{H^+},m_H)$ & 1726 &  1799 &  1174 &1546  \\
\hline
\end{tabular}
\caption
{MSSM masses in~GeV for a few sample parameter points 
for the $1+200$ model.}
\label{tab200}
\end{table}

\section{ Direct \& Indirect Detection rates 
for the $(1+200)$ Model:}
\label{detectionsfor1plus200}
Fig.\ref{xsection200A} shows the spin-independent scattering cross-section 
of the DM on nucleon, which gives the direct detection signal. The 
shaded (red) region covers the two WMAP relic density satisfying branches of 
Fig.\ref{200-spectrum}. Most of this region corresponds to the upper branch 
of Fig.\ref{200-spectrum}, except for the lower edge, which
corresponds to the lower branch. There is a thin line separating the two 
regions, which is barely visible in this figure.
 Note that the spin-independent
scattering is dominated by the Higgs exchange, whose coupling to the 
neutralino DM is proportional to the product of its higgsino and gaugino 
components. Thanks to the presence of $10-20\%$ of higgsino component in 
this model (Fig.\ref{200-spectrum}), the spin-independent cross-section is 
large enough to be detectable at the proposed superCDMS experiment\cite{supercdmssno}. 
Fig.\ref{xsection200B} shows the model prediction for the 
spin-dependent cross-section of the DM on nucleon. This is dominated by the 
$Z$ exchange whose coupling to the neutralino DM is proportional to the 
square of its higgsino component. The shaded (red) region 
covers the two WMAP relic 
density 
satisfying branches of Fig.\ref{200-spectrum}. Again most of this region 
corresponds to the upper branch, with only the lower edge corresponding to 
the lower branch. 
As in the case of the previous model, the spin-dependent cross-sections 
here is too small to be seen at direct detection 
experiments\cite{directdetexpt}.

\begin{figure}[!htb]
\vspace*{-0.05in}
\mygraph{xsection200A}{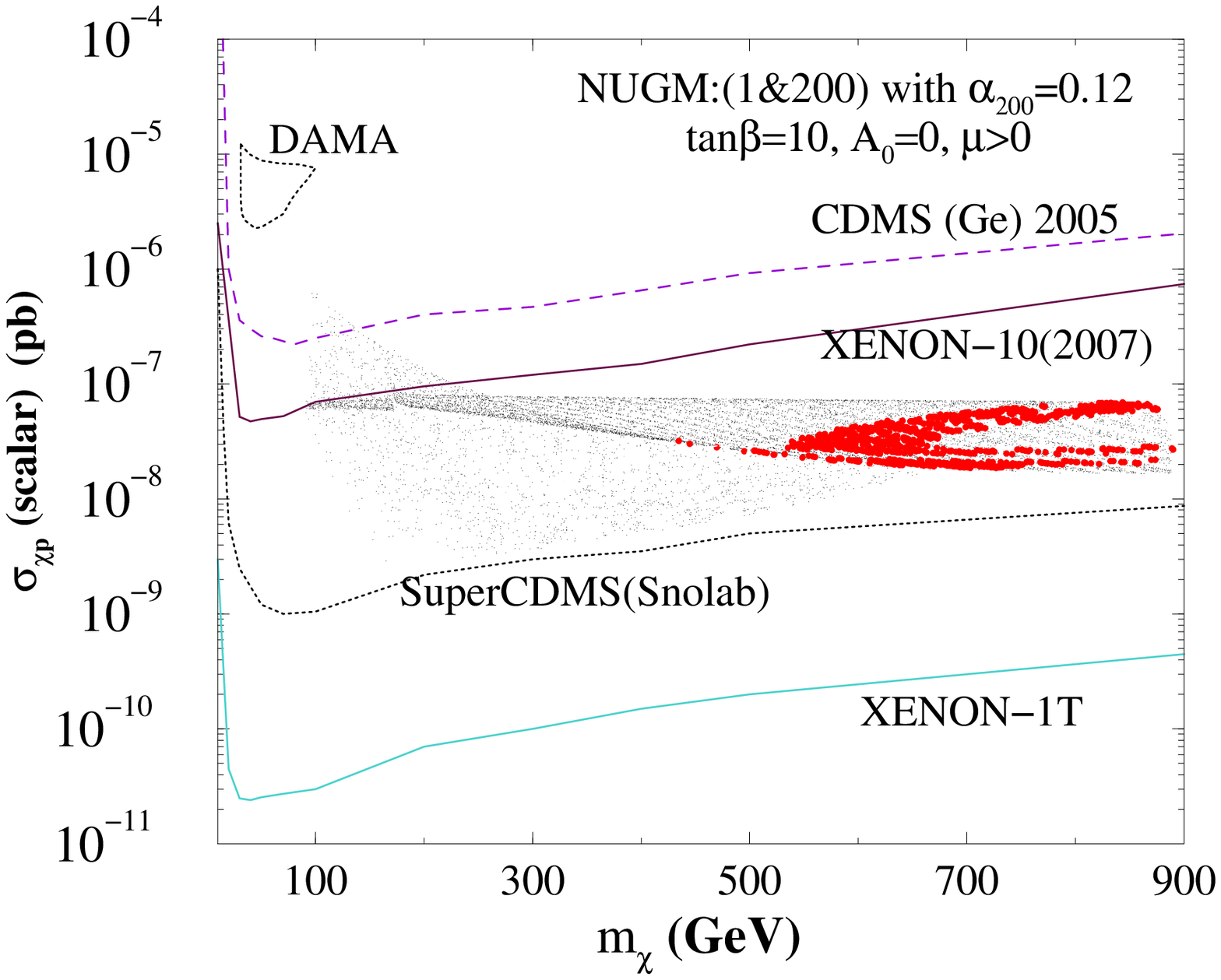}
\hspace*{0.5in}
\mygraph{xsection200B}{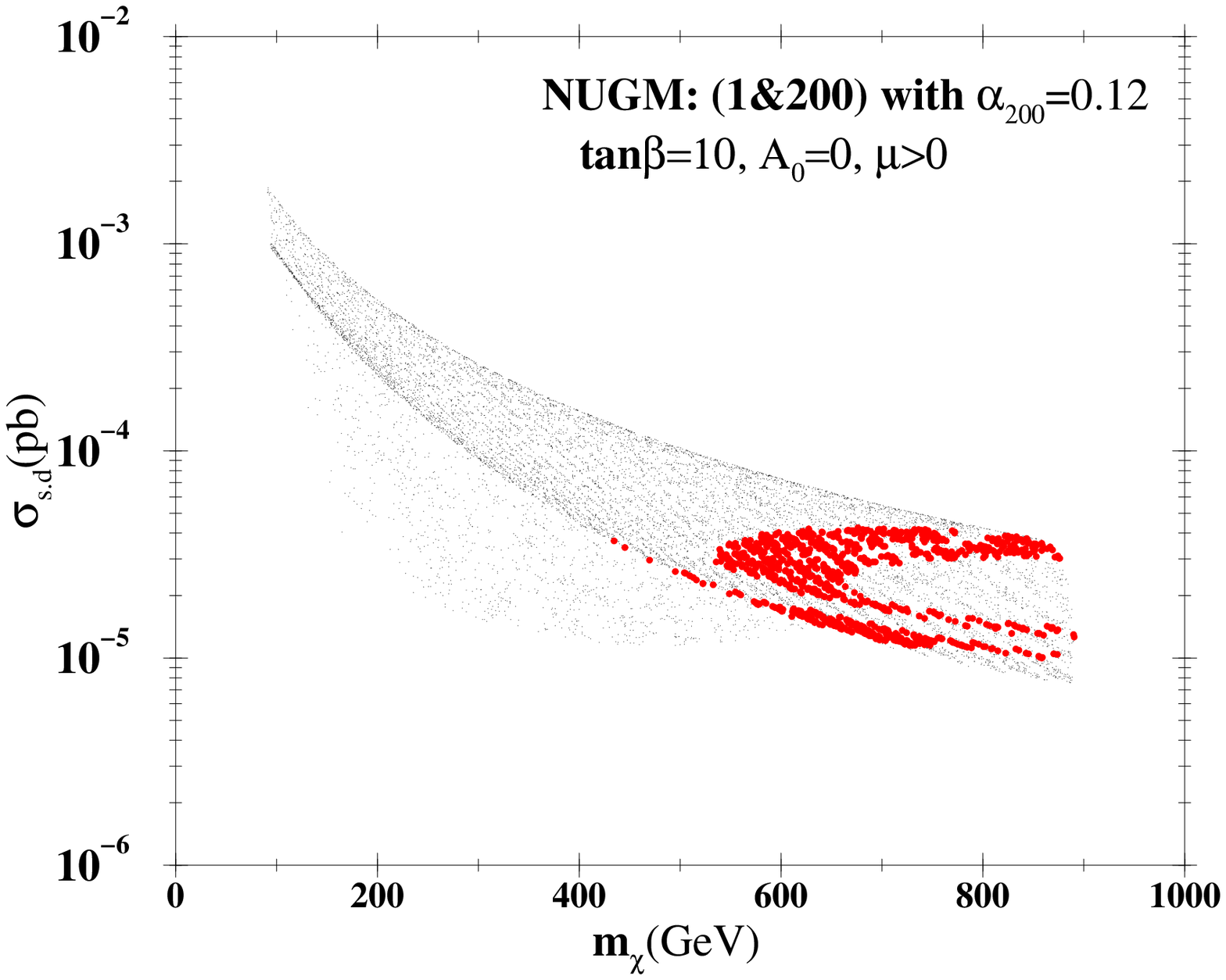}
\caption{ 
(a): Spin-independent neutralino-proton cross section for 
the $(1+200)$ model vs. LSP mass for the parameters shown in 
Fig.\ref{200-spectrum}. The WMAP satisfied regions are shown in red. 
Various limits from the recent (DAMA, CDMS and XENON-10) and 
the future (SuperCDMS and XENON-1T) experiments are shown. 
(b): Spin-dependent neutralino-proton cross section for 
the $(1+200)$ model vs. LSP mass for the parameters shown in 
Fig.\ref{200-spectrum}. The WMAP satisfied regions are shown in red. 
}
\label{xsection200}
\end{figure}

\begin{figure}[!h]
\centering
\includegraphics[width=0.7\textwidth,height=0.5\textwidth]
{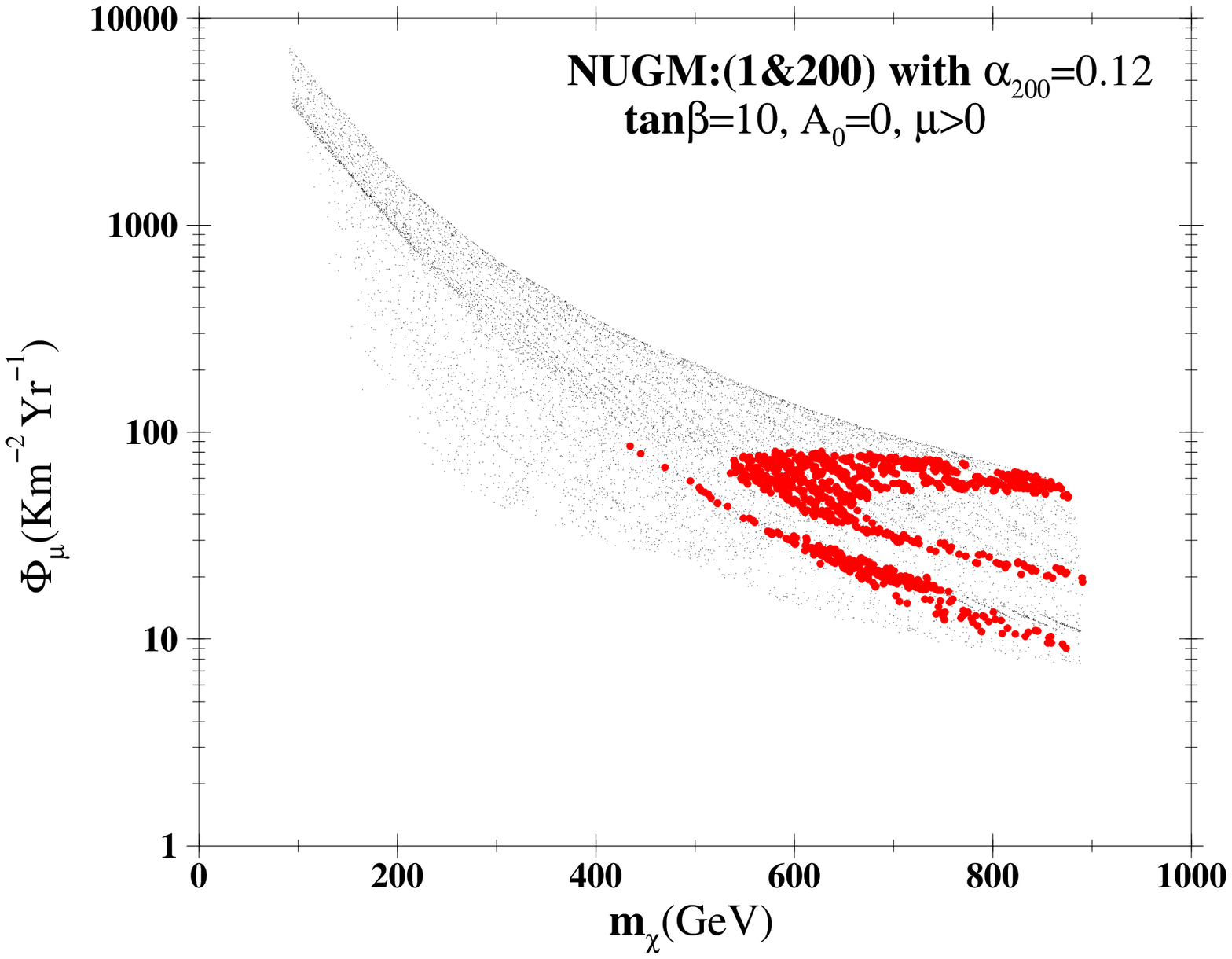}
\caption{
Neutralino annihilation induced muon flux from the Sun 
in ${km}^{-2} {yr}^{-1}$ vs 
LSP mass for the $(1+200)$ model for the parameters shown in
Fig.\ref{200-spectrum}. The WMAP satisfied regions are shown in red.
}
\label{muflux200}
\end{figure}

Similar to what was discussed in 
Section~\ref{detectionsfor1plus75} a larger Higgsino 
content of the LSP for $1+200$ scenario 
makes indirect detection of high energy neutrinos interesting. 
Fig.\ref{muflux200} shows the rate of 
the resulting muon events for the IceCube experiment. The upper and lower 
shaded (red) branches correspond to the respective WMAP satisfying branches 
of Fig.\ref{200-spectrum}, with a clearly visible separation between them. 
As in the previous model, one expects a detectable signal rate $\geq 10$ 
events per year at the IceCube experiment.

\begin{figure}[!htb]
\vspace*{-0.05in}
\mygraph{line200}{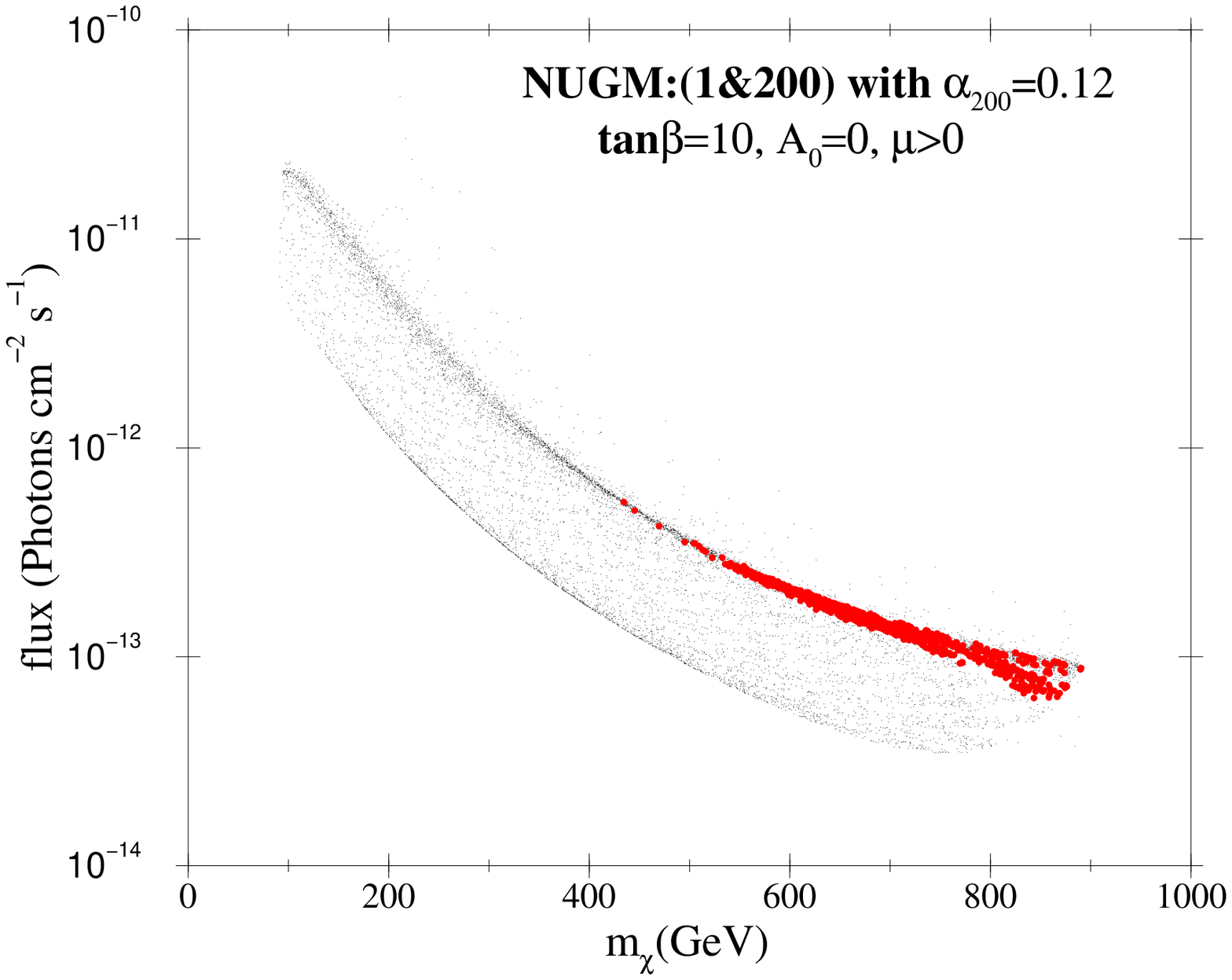}
\hspace*{0.5in}
\mygraph{cont200}{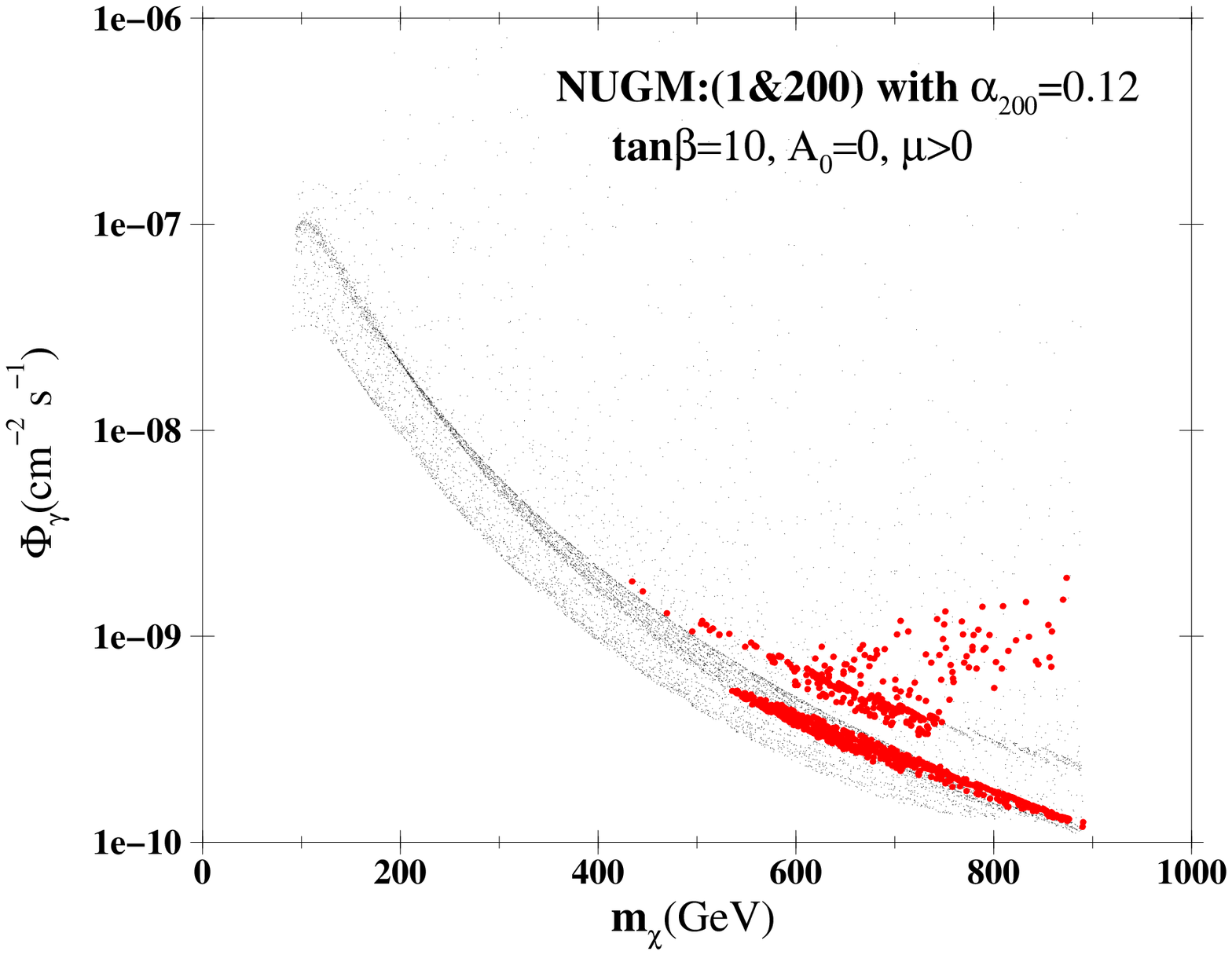}

\vspace*{0.5cm}
\hspace*{1.5in}
\mygraph{200phenergy}{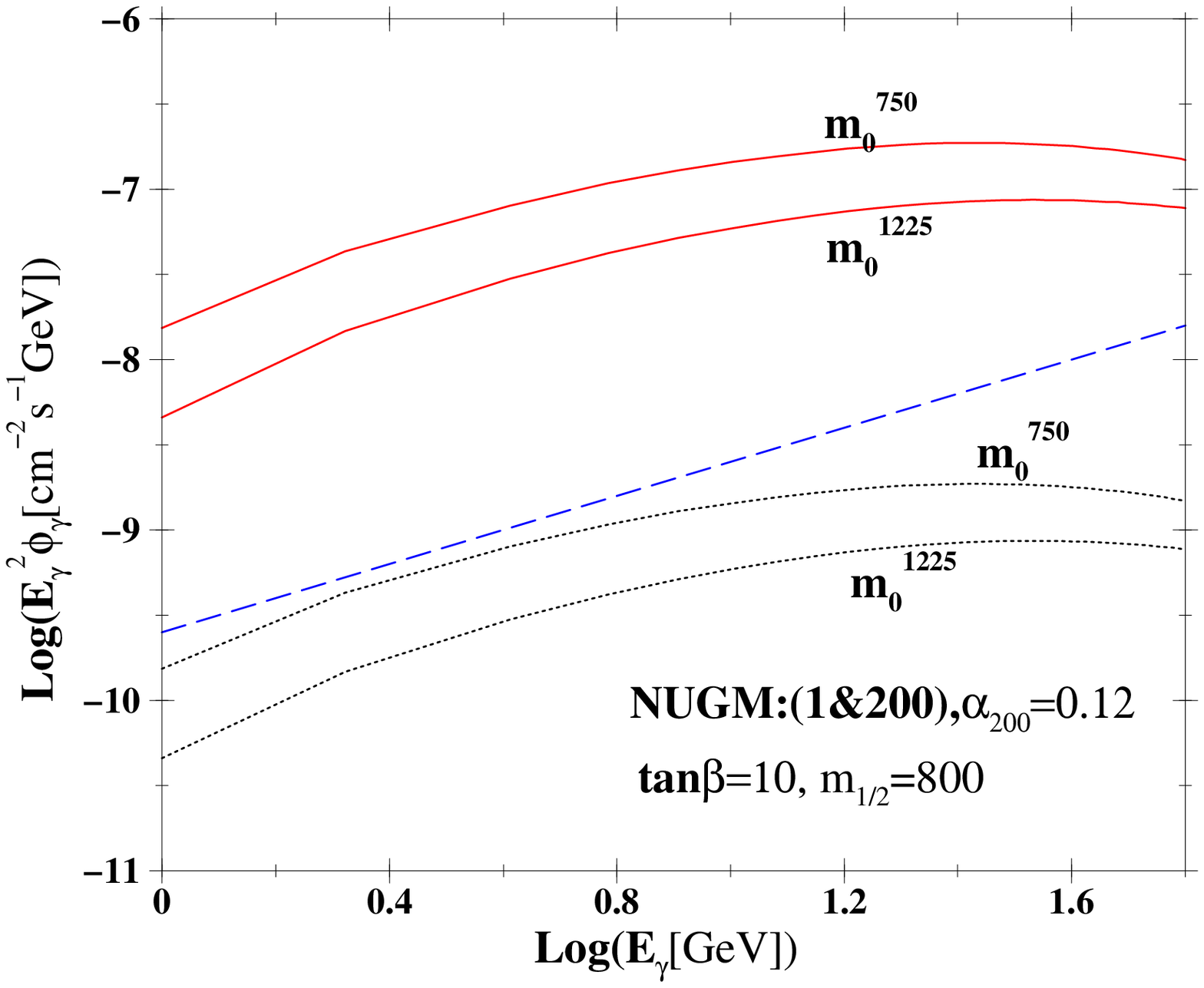}

\caption{
(a):Monochromatic $\gamma$-ray flux from DM 
pair annihilation near the galactic center 
vs. the LSP mass 
shown for the NFW profile of DM halo distribution and an aperture size of 
$\Delta\Omega = 10^{-3}$~sr for the $(1+200)$ Model. 
The parameters are same as in 
Fig.\ref{200-spectrum}. The WMAP satisfied regions are shown in red.
(b): Same as (a) except that the vertical axis refers to a continuous 
$\gamma$-ray flux above a threshold energy of $1$ GeV. (c):
Photon spectra expected from two particular LSPs corresponding to 
$m_0=750$~GeV and $m_0=1225$~GeV for the $(1+200)$ Model with 
$\mhalf=800$~GeV for an NFW profile. 
The other parameters are same as in Fig.\ref{200-spectrum}. 
The upper set of curves corresponds to a boost\cite{boostref} 
factor of $10^2$. The 
discovery limit of FGST/GLAST is shown as a blue straight line.  
}
\label{200-photon}
\end{figure}

Fig.\ref{200-photon} shows the rates of the line and the contunuum 
$\gamma$-ray signal coming from DM pair annihilation at the galactic core  
assuming the NFW profile 
of DM distribution\cite{{NFW}}. Fig.\ref{line200} shows the signal rate of the 
line $\gamma$-rays coming from DM pair annihilation,$\tilde \chi
 \tilde \chi \rightarrow \gamma \gamma (\gamma Z)$ via $W$ boson loop. 
Since the wino has a large Isospin gauge coupling, the resulting $\gamma$ ray 
signal here
is at least an order of magnitude larger than the previous model. 
The shaded (red) band corresponding to the WMAP satisfying region of this 
model, predicts a line $\gamma$-ray flux of $\sim 10^{-13}~\rm cm^{-2} ~s^{-1}
$. 
Several atmospheric Cerenkov experiments have reported $\gamma$-ray 
events\cite{magic,veritas,cangaroo,hess} at higher rate than this, but with 
continuous power-law energy spectrum, typical of supernova remnants. 
Therefore it will be hard to separate the DM signal from this 
background with the present energy and angular resolution of the ACT experiments.

Fig.\ref{cont200} shows the signal rate of the continuum $\gamma$-rays, 
coming from the tree-level pair annihilation processes of Eqs.\ref{75-up} 
and \ref{75-down},
with red dots corresponding to the WMAP satisfying regions. 
Since the most 
copious source of these $\gamma$ rays is the $b$-quark jet from Eq.
\ref{75-down}, the upper red dots correspond to the region of 
Fig.\ref{200-spectrum} with large resonant annihilation contribution. 
Finally  Fig.\ref{200phenergy} shows the continuum $\gamma$-ray spectrum 
for some representative points of this model, along with the discovery 
limit of the FGST/GLAST experiment\cite{gelmini08}. It shows that a boost factor of 
$\sim 10^2$ to see this $\gamma$-ray signal in FGST/GLAST for the assumed NFW 
profile of DM distribution.

\section{ LHC Signatures of the $(1+75)$ and $(1+200)$ Models:}

A quantitative investigation of the LHC signature of these two nonuniversal 
gaugino mass models is beyond the scope of this work. 
We shall only remark on some 
prominent features, which are quite evident from the SUSY spectra 
of Table~\ref{tab75} and \ref{tab200}. Table~\ref{tab75} clearly shows an 
inverted mass hierarchy for the $(1+75)$ model where the lighter stop 
is about $30\%$ lighter than the 1st and 2nd generation squarks. Moreover 
it is lighter than the gluinos over most of the parameter space. So one 
expects a SUSY signal from the channels 
\beqn
\label{lhc}
pp \rightarrow \tilde g \tilde g \rightarrow  t t \tilde t_1 \tilde t_1
,\ \ \ pp \rightarrow  \tilde t_1 \tilde t_1
\eeqn 
inclusive of their antiparticles. In view of the singlet dominance of 
$\tilde t_1$, one expects it to decay via mainly gauge (Yukawa) coupling to
$\tilde B (\tilde H)$ into
\beqn
\label{decay-75}
\tilde t_1 \rightarrow t \chi^0_{1,2,3}, \ \ \ b {\tilde \chi}^{+}_1.
\eeqn
Since all these neutralino and chargino states are nearly degenerate, 
we expect their $p_T$ to be largely carried by the LSP($\chi^0_{1}$), 
giving relatively 
large missing-$p_T$ ($\ptslash$) events along with soft jets (leptons). 
Note however that one expects hard leptons and $b$-jets from the decay of 
the top quarks in (\ref{lhc}) and (\ref{decay-75}), resulting in hard and 
isolated multilepton$+$missing $p_T$ signal accompanied by 
$4(2)$ $b$-jets for the pair production of gluon (stop). 

The SUSY spectra of Table~\ref{tab200} also shows an inverted hierarchy 
for the $(1+200)$ model with the $\tilde t_1$ being much lighter than the 
other squarks. Moreover it is lighter than the gluino over essentially 
the entire parameter space. Because of the rather heavy mass range of 
squarks and gluino in this model we expect a very important contribution to 
the LHC SUSY signal to come from the pair production of $\tilde t_1$. 
Note that in this case the dominant decay channels are 
\beqn
\label{decay-200}
\tilde t_1 \rightarrow t \chi^0_{1,2,3,4}, \ \ \ b {\tilde \chi}^{+}_2,
\eeqn 
since the higgsino dominated states are the heavier chargino and 
neutralino states 
$({\tilde \chi}^{+}_2, \chi^0_{3,4})$. However these states are also quite 
close to the LSP $\chi^0_{1}$ in this case. Therefore one expects signal 
characteristics as the earlier model. It should be mentioned here that 
similar characteristics of the LHC signal was noted for the focus point 
region of the mSUGRA model in \cite{focusLHC1}, and has since been studied in 
detail in \cite{focusLHC2}. One expects similar features for these nonuniversal 
gaugino mass models of the mixed neutralino LSP as well. Only in these 
models one can have relatively light scalar mass $m_0$, so that the 
lighter $\tilde t_1$ is seen to be even lighter than the gluino, unlike the 
focus point case.

\section{Conclusion}
We have investigated the dark matter properties of non-universal gaugino mass 
scenarios where the gauge kinetic energy function is a mixture 
of chiral superfields that transforms as singlet and the 75-dimensional or 
the singlet and the 200 dimensional representation of SU(5).  
The mixing of the 
representaions are chosen 
so as to probe the dark matter annihilation properties of LSPs that 
are either a strong mixture of bino and Higgsinos in the 1+75 model  or 
a mixture of bino-wino or even a mixture of bino-wino-higgsinos 
in the 1+200 model. For each of 
the scenarios we have analyzed the dark matter relic density in relation to 
the WMAP data and we have obtained broad regions of parameter space in the 
$m_0-\mhalf$ plane that satisfy the WMAP data. 
We have identified the annihilation and coannihilation 
channels. We have further computed the direct detection 
rates of neutralino DM for 
the spin-independent and spin-dependent $\chi_1^0-p$ cross sections. The 
spin-independent cross section ranges would be successfully 
probed in the near future via XENON-1T. We have also computed the 
neutralino-annihilation induced muon flux from the Sun that originates 
from the high energy neutrinos produced via pair-annihilation of LSPs in the 
solar core. We have found promising signals that will 
easily be probed in the IceCube experiment. We have also obtained interesting 
photon signal rates (both monochromatic and continuum) for the 
indirect detection of LSPs via $\gamma$-rays. FGST/GLAST will be able to probe 
a significant region of parameter space with a possible  
boost factor. Finally, we have described a few 
distinctive features of the expected LHC signals of these models.

\noindent
{\bf Acknowledgments}\\ 
This work was initiated during the NORDITA Summer Programme (2008) 
at Stockholm. 
UC and DPR thank the organizers of this programme for their kind hospitality.
The work of DPR was partially supported by 
BRNS (DAE) under the Raja Ramanna Fellowship Scheme. DD would like 
to thank the Council of Scientific and Industrial Research, Govt. of India 
for financial support.

\end{document}